# The Visible and Near Infrared module of EChO

A. Adriani[(1)], G. Bellucci[(1)], L. Gambicorti[(2,5)], M. Focardi[(2)], E. Oliva[(2)], M. Farina[(1)], A.M. Di Giorgio[(1)], F. Santoli[(1)], E. Pace[(3)], G. Piccioni[(1)], G. Filacchione[(1)], M. Pancrazzi[(2)], A. Tozzi[(2)], G. Micela[(4)].

1) INAF Istituto di Astrofisica e Planetologia Spaziali, Via del Fosso del Cavaliere, 00133 Roma, ITALY
2) INAF Osservatorio Astrofisico di Arcetri, Largo E. Fermi 5, 50125 Firenze, ITALY
3) Dipartimento di Fisica e Astronomia, Università degli Studi di Firenze, Largo E. Fermi 2, 50125 Firenze, ITALY
4) INAF Osservatorio Astronomico di Palermo, Piazza del Parlamento 1, 90134 Palermo, ITALY
5) Physics Institute Space Research and Planetary Sciences, Sidlerstrasse 5, CH 3012 Bern, Switzerland (current address)

Corresponding author: Giancarlo Bellucci
Email: giancarlo.bellucci@iaps.inaf.it
Tel: +390645488361
Fax: +390645488070

**Abstract**

The Visible and Near Infrared (VNIR) is one of the modules of EChO, the Exoplanets Characterization Observatory proposed to ESA for an M-class mission. EChO is aimed to observe planets while transiting by their suns. Then the instrument had to be designed to assure a high efficiency over the whole spectral range. In fact, it has to be able to observe stars with an apparent magnitude $M_v = 9 \div 12$ and to see contrasts of the order of $10^{-4} \div 10^{-5}$ necessary to reveal the characteristics of the atmospheres of the exoplanets under investigation.

VNIR is a spectrometer in a cross-dispersed configuration, covering the $0.4 \div 2.5$ μm spectral range with a resolving power of about 330 and a field of view of 2 arcsec. It is functionally split into two channels respectively working in the $0.4 \div 1.0$ μm and $1.0 \div 2.5$ μm spectral ranges. Such a solution is imposed by the fact the light at short wavelengths has to be shared with the EChO Fine Guiding System (FGS) devoted to the pointing of the stars under observation.

The spectrometer makes use of a HgCdTe detector of 512 by 512 pixels, 18 μm pitch and working at a temperature of 45K as the entire VNIR optical bench. The instrument has been interfaced to the telescope optics by two optical fibers, one per channel, to assure an easier coupling and an easier colocation of the instrument inside the EChO optical bench.

1. Introduction

The discovery of over a thousand exoplanets has revealed an unexpectedly diverse planet population. We see gas giants in few-day orbits, whole multi-planet systems within the orbit of Mercury, and new populations of planets with masses between that of the Earth and Neptune – all unknown in the Solar System. Observations to date have shown that our Solar System is certainly not representative of the general population of planets in our Milky Way (www.exoplanet.eu). The key science questions that urgently need addressing by EChO are therefore: What are exoplanets made of? Why are planets as they are? How do planetary systems work and what causes the exceptional diversity observed as compared to the Solar System? The EChO mission (Tinetti et al. 2012) will take up the challenge to explain this diversity in terms of formation, evolution, internal structure and planet and atmospheric composition. This requires in-depth spectroscopic knowledge of the atmospheres of a large and well-defined planet sample for which precise physical, chemical and dynamical information can be obtained.

In order to fulfill this ambitious scientific programme, EChO is designed as a dedicated survey mission for transit and eclipse spectroscopy capable of observing a large, diverse and well-defined planet sample within its four-year mission lifetime. The transit and eclipse spectroscopy method, whereby the signal from the star and planet are differentiated using knowledge of the planetary ephemerides, allows us to measure atmospheric signals from the planet at flux levels of at least $10^{-4}$ relative to the star. This can only be achieved in conjunction with a carefully designed stable payload and satellite platform. It is also necessary to provide an instantaneous broad-wavelength coverage to detect as many molecular species as possible, to probe the thermal structure of the planetary atmospheres and to correct for the contaminating effects of the stellar photosphere. This requires wavelength coverage of at least 0.55 to 11 μm with a goal of covering from 0.4 to 16 μm. Only modest spectral resolving power is needed, with R~300 for wavelengths less than 5 μm and R~30 for wavelengths greater than this. The transit spectroscopy technique means that no spatial resolution is required. A telescope collecting area of about 1 $m^2$ is sufficiently large to achieve the necessary spectro-photometric precision: in practice the telescope will be 1.13 $m^2$, diffraction limited at 3 μm. Placing the satellite at L2 provides a cold and stable thermal environment as well as a large field of regard to allow efficient time-critical observation of targets randomly distributed over the sky. EChO is designed, without compromise, to achieve a single goal: exoplanet spectroscopy. The spectral coverage and signal-to-noise ratio to be achieved by EChO, thanks to its high stability and dedicated design, will be a game changer by allowing atmospheric compositions to be measured with unparalleled exactness: at least a factor 10 more precise and a factor 10 to 1000 more accurate than current observations. This will enable the detection of molecular abundances three orders of magnitude lower than currently possible. Combining these data with estimates of planetary bulk compositions from accurate measurements of their radii and masses will allow degeneracies associated with planetary interior modeling to be broken, giving unique insight into the interior structure and elemental abundances of these alien worlds.

EChO will carry a single, high stability, spectrometer instrument. The baseline instrument for EChO is a modular, three-channel, highly integrated, common field of view, spectrometer that covers the full EChO required wavelength range of 0.55 μm to 11.0 μm. The baseline design includes the goal wavelength extension to 0.4 μm while an optional LWIR channel extends the range to the goal wavelength of 16.0 μm. Also included in the payload instrument is the Fine Guidance System (FGS), necessary to provide closed-loop feedback to the high stability spacecraft pointing. The required spectral resolving powers of 300 or 30 are achieved or exceeded throughout the band. The baseline design largely uses technologies with a high degree of technical maturity.

The spectrometer channels share a common field of view, with the spectral division achieved using a dichroic chain operating in long-pass mode. The core science channels are a cross-dispersed spectrometer VNIR module covering from 0.4 to ~2.5 μm, a grism spectrometer SWIR module covering from 2.5 to 5.3 μm, and a prism spectrometer MWIR module covering from 5.3 to 11 μm. All science modules and the FGS are accommodated on a common Instrument Optical Bench. The payload instrumentation operates passively cooled at ~45K with a dedicated instrument radiator for cooling the FGS, VNIR and SWIR detectors to 40 K. An Active Cooler System based on a Neon Joule-Thomson Cooler provides the additional cooling to ~28 K which is required for the longer wavelength channels.

In the following, the characteristics of the VNIR module are described in detail.

## 2. Module Design
### 2.1. Optical Layout

The system covers the spectral range between 0.4 and 2.5 µm without gaps and the resulting resolving power is nearly constant, R≈330. The wide spectral range is achieved through the combined use of a grating with a ruling of 14.3 grooves/mm and blaze angle of 3.3° for wavelength dispersion in horizontal direction and an order sorting calcium fluoride prism (angle 22°), which separates the orders along the vertical direction. The collimator (M1) and the prism are used in double pass (see figure 1). The prism is the only optical element used in transmission. All other optics are made of reflecting surfaces: 2 off-axis conic mirrors, 1 spherical mirror, 1 flat mirror and 1 grating. All reflecting elements will be made of the same aluminium alloy as the optical bench. This simplifies the mechanical mount and alignment of the system. The light is fed to the spectrometer via two fibres positioned on the side of the M2 mirror. The fibres are commercial fused-silica with ultra-low OH content and core diameter of 50 µm. The fibres are separately fed by two identical off-axis parabolic mirrors (M0) which intercept the collimated light transmitted from the first dichroic (D1b), IR, and reflected by the beam-splitter, VIS. The use of an optical fibres coupling gives a larger flexibility in the location of the VNIR spectrometer within the EChO payload module. The VNIR characteristics are summarized in Table 1.

A Mercury Cadmium Telluride (MCT, HgCdTe) detector has been considered for VNIR (its technical characteristics are detailed later in section 4). Figure 2 shows the observable spectral orders, m, projected on the MCT array, starting from m = 3 at the bottom (near infrared spectral range) to m = 20 on the top (visual spectral range).

Namely, the figure shows the distribution of the light on the array between 2500 nm ($m = 3$) and 400 nm ($m = 20$). The central wavelength in each order $m$, positioned at the blaze angle of the grating, is given by the relationship $\lambda = 8.1/m$ µm. The VIS and IR spectral ranges are separated on the detector because the fibres are placed at the spectrometer entrance are separated by 1 mm. In general, most wavelengths are sampled twice on different orders, i.e. in different areas of the detector, as shown in figure 2. The spectrum in each order is spread across several pixels in the vertical direction. Thus, a sum over 5 pixels will be done to increase the sensitivity of the system in order to provide a so-called spectral channel. The last two instrumental features, about wavelength sampling, also have the advantage to reduce systematic errors in the measurements.

As previously said, the coupling of the VNIR module to the telescope will be done through the use of a dichroic element that will select and direct the visible and near infrared light towards the combined system VNIR and FGS. A beam-splitter is foreseen to further divide the light beam between FGS and VNIR. The balance of this beam-splitter will need to be studied in conjunction with the FGS team during the assessment phase to maximize the science return while maintaining sufficient signal for the guider system. As the performance of the module optics should be very good to assure the observations of transient planets in transit or in occultation of a star, the detector is going to be a key element in the system. In order to meet the EChO visible channel performance requirements, it is possible to pursue different ways, based on different detectors and readout electronics as well as on the optical spectrometer design characteristics.

| *Parameter* | *Value* |
|---|---|
| Spectral Range | 0.4÷2.5 µm |
| Resolving power | ≈ 330 |
| FOV | 2 arcsec |
| SNR | 25 @1.5 µm for 1 sec exposure, star Mv = 9 |
| Detector Type | HgCdTe |
| Detector size | 512 x 512 pixels |
| Pixel size | 18 µm square |
| Pixel binning | 5x5 |
| Signal digitalization | 16 bits |
| Working Temperature | 40÷45K |

Table 1

### 2.1.1. Internal Calibration Unit

The instrument calibration is going to be performed looking at a known reference star before and after any target observation. The star calibration is meant to verify the position of the spectral lines However, it is important to monitor the stability of the instrument and, in particular, of the detector during each observation session. The observation session is supposed to vary from minutes to about 10 hours depending on the characteristic of the target itself. In order to assure the quality of the measurements the calibration unit has to guarantee the possibility of monitoring possible instrumental variations of the order and better of $10^{-4}$. The calibration unit will be equipped with two Halogen-Tungsten lamps for redundancy. Those kinds of lamps are currently used as spectral calibration sources of optical systems and are the baseline for the development of the VNIR calibration unit. The calibration lamps will be equipped with a close loop control system to assure the requested stability over the observation time. The lamps will have color temperature higher that 3000K and it will be operating for very short times during the observation sessions.

The lamps inject their light into an integration sphere, which will have two output fibers that will feed the two input fibers to the spectrometer (ranges 0.4÷1.0 µm and 1.0÷2.5 µm respectively). Figure 3 gives the spectrum in input to the fibers. The feeding of the main fibers will be done using 2in-1out fiber connectors. The two fibers will be illuminated at the same time.

The calibration unit will be located in a separate box on a side of the service box where the mirrors collect the light from the VNIR feeding optics and focus it on the optical fibers inputs. Figure 4 shows the calibration unit and its arrangement on the service side of the VNIR optical bench.

### 2.2. Mechanical and Thermal design

VNIR instrument is housed in a mechanical structure, connected to the spacecraft interface through an isostatic mount. The optical elements (mirrors and prism) are shown in the right panel of figure 5. The box will be built in a way to keep the weight as small as possible and the material will be the same of the payload optical bench. The figure shows the box without the calibration unit mounted below the spectrometer optical bench. The overall dimensions are: 342 x 325 x 190 mm and the dimensions are depicted in the lower left panel. A view inside the box is given in the right panels of the figure where the location of the optical elements of the spectrometer is shown. The lower part of the VNIR optical bench will be dedicated to the services to spectrometer: the input box where the mirrors concentrate the light on the optical fibers and the calibration unit in two separated box in order to minimize light and thermal contamination of the rest of the instrument. The VNIR calibration unit switches on/off and overall control will be performed by the EChO Instrument Control Unit (ICU) (Focardi et al. 2014).

The mass of the instrument is estimated to be about 6.8 kg (20% margin included). The VNIR First Resonant Frequency is planned to be larger than 150 Hz. More accurate estimation can be done by a structural analysis that will take into account the effective vibration levels of the spacecraft during the launch.

The instrument will be built in the same material of the payload optical bench and the box will be thermally linked through its feet to it. The optical box and the VNIR CFEE (Cold Front End Electronics – SIDECAR ASIC as baseline) are supposed to be at temperatures lower than 50K. The CFEE would be located on the telescope optical bench. The detector is planned to work at a temperature in a range of 40÷45K, dissipating about 30 mW.

### 3. Instrument Performances

The grating's orders of diffraction, as shown in figure 2, on the detector would not be equally illuminated if the input light would have a constant intensity over the entire spectrum because the grating's efficiency changes along the order. The maximum efficiency is around the center of the blue curves in figure 2. In this spectrometer configuration some wavelengths can be observed on two adjacent diffraction orders. To completely recover the light at those wavelengths the signal coming from the adjacent order has to be summed. The sum has to be done to maximize the result and keep the highest feasible signal to noise ratio. A reasonable compromise has been found in summing the adjacent orders when the grating efficiency is higher than 80% with respect to the maximum. The result is a component of the Instrument Transfer Function (ITF) of the instrument that will be given as result of the on-ground instrumental calibrations by measuring and combining the optical efficiency of the spectrometer and the detector performances. Figure 6 shows the spectrometer efficiency calculated with 80% criterion. The present calculation has been done by considering aluminum mirrors without any coating to improve the performances at wavelengths lower than 1.0 µm. In the picture the expected behavior obtained by the use of coated aluminum or protected silver mirrors is also shown for comparison.

The coupling of the light from the telescope pupil to the input of the optical fiber is affected by the effectiveness of the telescope pointing system (FGS, Fine Guidance System). The effect of the mean performance error (MPE) is a loss of efficiency from observation to observation. The effect of the relative performance error (RPE) is a photometric error within an observation. A simulation is performed at two representative wavelengths (0.8 and 2.5µm) to estimate the magnitude of both effects. The illumination pattern of the telescope is obtained from optical modeling. The energy collected by the fiber is then studied as a function of MPE, RPE and PRE (performance reproducibility error). The MPE is varied in accordance to EIDA-R-0470. The impact of three different RPEs is studied: i) RPE1 = 30 mas-rms from 1 to 10Hz; ii) RPE2 = 50 mas-rms from 1 to 300Hz; iii) RPE3 = 130 mas-rms from 1 to 300Hz. These three cases correspond to three different AOCS (Attitude and Orbit Control Systems) solutions. A fixed PRE = 20 mas-rms from 0.020 to 4mHz is used in this simulation.

The results of the simulations are discussed in Waldmann and Pascale 2013 and are here briefly summarized. The effect of the MPE on the normalized transmitted energy is shown in figure 7. The combined effect of the RPE and PRE on the photometric error is shown in figure 8. The worst case photometric error is obtained when observing a bright target (a star with visual magnitude Mv = 4) with the RPE3 option and results in 10% of the total allowed system noise variance in one second of integration for this channel.

The analysed optical system is the Echo Telescope and the concentrating system (f#4) in input of fibre. The configuration optimized consists in primary mirror telescope distance $M_{T1}$-$M_{T2}$=1.500 mm, the configuration defocused determines WFE 250 rms with shift M1-M2 position of 87 micron. (WFE calculated at 1 micron wavelength). The fibre with 50-µm diameter corresponds to a Field of View (FOV) of 2 arcsec.

The spot diagram and the encircled energy are simulated to verify the requirement. Simulation of encircled energy in 250rms WFE generated by defocusing ($M_{T1}$ back with respect to $M_{T2}$) of 0.087 mm (FOV/2= 1 arcsec ($2.7 \times 10^{-5}$ deg). (The simulation considers only defocusing shift on optical axis. It is not a complete evaluation of efficiency because the tilt and lateral shift are not included).

Figure 9 shows that the spot diagram of the aberrated beam after defocusing is collected inside the fiber diameter.

Table 2 resumes all obtained results, spot diagram and Encircled Energy collected on entrance fiber of VNIR channel.

| WFE rms @1 micron | 7.3 rms | 250 rms |
|---|---|---|
| Defocus shift M1-M2 (micron) | 0 | 87 |
| GEO radius spot (micron) | Airy radius 4.9 | 12.47 |
| Encircled Energy in-fibre diameter (2arcsec) | 96.75% | 96.75% |

Table 2

Spot diagram inside the fibre diameter and collected Encircled Energy (96.75%) demonstrate that the introduction of a defocusing of 250 WFE rms in entrance beam of fibre.

The efficiency of a fiber is the product of three effects, namely internal transmission (which is at most 95% in our case), reflection losses at the entrance/exit (which amount to 6%) and focal ratio degradation (FRD), which measures the fraction of light exiting from the fiber within a given solid angle. The value of FRD depends on the aperture angle (i.e. the focal aperture F/#) by which the fiber is fed, and by the focal aperture accepted by the spectrometer. The VNIR fiber receives an F/4 input beam and feeds the spectrometer with an F/3.5 output beam. Therefore, the FRD losses are about 5% and total efficiency is about 85%.

The light from the telescope can be fed to the fiber on the image plane or on the pupil plane. The former solution is used in HARPS, the ultra-high precision astronomical spectrometer which has reached the highest accuracy in the detection of extra-solar planets. On the other hand, pupil-feeding are often used in fiber-fed astronomical instruments. In the case of VNIR we can use both solutions, the only difference being the curvature of the input surface of the fiber, which is flat in case of image-feeding. For pupil-feeding, instead, the curvature is such that the first part of the fiber acts as a micro-lens adapter. We plan to test both solutions and select the one providing the best performances in terms of total efficiency and scrambling gain.

4. **The detector choice**

For the visible and near infrared channel, two options have been considered for VNIR: 512 x 512 matrix with 18 μm square pixels and 256x256 format with 30 μm square pixels Mercury Cadmium Telluride, MCT, operating at high frame rate (of the order of 10Hz). The first option is considered as baseline in this paper. MCTs have a good efficiency in the VNIR spectral range keeping a very low readout noise. Like other spectrometric EChO channels working in the infrared, the choice of an MCT permits the detector to work at a temperature around 40÷45 K, matching that of the optical bench of the modules. This fact will allow the instrument to have a very low thermal noise. From the performance point of view, readout noise, pixel size and dark current are the most crucial parameters that have been taken into consideration for the selection. This because the VNIR signal to noise ratio drops below 1 micron and detector noise performances are crucial to meet the requirements. As far as the 512x512 format is concerned, Selex and US manufacturers (Teledyne and Raytheon) offer comparable performances. For example, Selex provides a readout noise <23 $e^-$ rms and a dark current of the order of $3 \times 10^{-2}$ electrons/second at 80K for a pixel pitch of 24 μm (it should be much lower than this at 40÷45 K and for a 15 μm pixel). Teledyne provides a readout noise <18 $e^-$ rms and a dark current of the order of $10^{-2}$ electrons/second at 77K. The Selex performances for the 256x256 format are a bit lower than the Teledyne one, providing a readout noise <65 $e^-$ rms. While the US detectors appears to be in a mature state, Selex has a series of technical activities ongoing and planning to improve the performances of the VNIR detector, taking one of these devices at TRL 5 at beginning of 2015. From both the technical and programmatic information we have received from manufacturers, we assume Teledyne as baseline and Selex as a backup. Teledyne detectors can also be connected directly to the SIDECAR ASIC, chosen as baseline for the VNIR CFEE (as well as for the SWIR and FGS modules). This solution is better in terms of power consumption, thermal coupling and simplifies the overall harnessing between the detector and the CFEE and between CFEE and WFEE/ICU.

5. **Noise effects studies**

The VNIR focal plane is made by MCT arrays. A study has been carried out to evaluate the best readout mode to adopt with these detectors, taking into account the following main aspects: the need to minimise the equivalent noise in both bright and faint stars observations, the need to detect and correct for the cosmic rays hits effects and, finally, the need to maintain the overall data rate within the allowed telemetry budget ($\leq$ 5 Gbits/day).

The MCT detectors allow for non-destructive readout modes, such that multiple readouts are possible without disturbing ongoing integration.

In Figure 10 a non destructive readout sampling scheme is shown, for a single MCT pixel, in which the detectors integrating ramps are indicated in blu. In the sample up-the-ramp readout mode, the detectors readouts are equally spaced in time, sampling uniformly the ramp. By collecting all samples it is possible to fit the ramp slope. Provided that the number of samples is statistically significant, in case of cosmic rays hits, a jump or even a smooth modification of the slope can be detected and the corresponding samples rejected. This method is accurate but quite demanding in terms of real time processing power. In the multi accumulate readout mode, only contigous groups of samples are considered. The groups are equally spaced in time, but the samples between the groups are discarded. In Figure 1 the samples groups are highlighted in red.

Multi-accumulate and sample Up-The-Ramp readouts can be considered as the building blocks of all non-destructive readout modes such as Correlated Double Sampling (CDS), Multiple CDS (MCDS, also known as Fowler-M), sample Up-The-Ramp (UTR), Multi-Accumulate (MACC) and Differential Multi-Accumulate (DMACC). Refer to Finger et al. 2005[2] and references therein for a more detailed description of all the modes.

All these modes reduce the equivalent readout noise increasing the signal to noise ratio. The general expression for the total noise variance of an electronically shuttered instrument using the non-destructive readout can be computed using well known relations based on fundamental principles. It has been presented for the first time in its complete form by Rauscher et al. 2007 [3] and is reported hereinbelow:

$$\sigma^2 = \frac{12(n-1)}{nm(n+1)} R^2 + \frac{6(n^2+1)}{5n(n+1)}(n-1)(t_g)f - \frac{2(2m^2-1)(n-1)}{mn(n+1)}(m-1)(t_f)f$$

$R$ is the readout noise and $f$ is the flux, including photon noise and dark currents. $R$ is in unit of $e^-$ rms per read and $f$ is in unit of $e^- s^{-1} spaxel^{-1}$ (Spaxels are pixels bins of $j \times k$ pixels, $j$ along the spectral direction and $k$ along the spatial direction); $n$ is the numbers of stores per exposure (groups), $m$ is the numbers of samples to coadd per group. The frame time $t_f$ is the time interval between reading pixel [0, 0] in one frame and reading the same pixel in the next frame within the same group (sampling frequency). The group time $t_g$ is the time interval between reading pixel [0, 0] in the first frame of one group and reading the same pixel in the first frame of the next group.

We used this relation to evaluate the noise expected for the VNIR detectors when read out using the sample up-the-ramp method. The result has then been compared with the system requirements for the two different detector arrays under study for the Echo mission, see Farina et al. 2013. The aim of the work has been to provide indications on how to optimize the EChO focal plane arrays sampling rate and data processing procedures in order to achieve the best signal to noise ratio and to identify and remove Cosmic rays effects. The results of this activity will also be used to dimension the on-board hardware and to define the architecture of the onboard data processing software.

The sampling rates of 8 Hz for bright sources and 1/16 Hz for faint sources have been considered; see Focardi et al. 2014 for a detailed description of the reasons behind the choice of these rates. The adopted integration times are 3s for bright sources and 600s for faint sources. Given the estimated input fluxes for the two kind of sources, these times allow to couple with the maximum detectors well capacity in both cases. All comparisons are made assuming an operating temperature of 40 K.

For bright sources only the Teledyne detectors provide an expected total noise below the scientific requirement. The Selex estimated noise is always well above the noise requirement and the obtained trend is not decreasing with the increase of the integration time.

In particular, in the case of the Teledyne detectors, for $m \geq 2$, the minimum $n$ to satisfy the requirement is always very low. This situation allows to tune the overall measurement duration (max integration time) based only on the deglitching procedure performances, keeping it as short as possible, thus minimizing the expected number of cosmic hits.

In case of faint sources the results obtained for the planned 1/16Hz sampling rate are similar to the previous ones, even if in this case the Selex detectors are able to meet the noise requirements in at least one case, with m=3 and the minimum $n$ equal to 7. More generally, at 1/16 Hz both the detectors can fall below the noise threshold with $m$=3 and $n$=12.

Considering these preliminary results Teledyne sensors better combine bright and faint sources results, while Selex detectors need to be better tested.

In the next mission phases this method will be validated by tests and more insights will be done. In particular, we are confident of reaching the noise requirement, for bright sources, using a sampling data rate lower than those reported previously and, for faint sources, to reach the noise threshold also with $m=1$ (pure up-the-ramp fitting, with no multi-accumulation). This because the maximum integration time compatible with the detectors well depth is much shorter than the time scale of sources variability; therefore many ramps could be coadded improving the signal to noise ratio.

With respect to the expected cosmic hits rate, assumptions based on studies made for the JWST telescope (see Fixsen et al. 2000, Rauscher et al. 2000) give an expected rate of cosmic events with impact on the detector confined between 5 and 30 events/s/cm$^2$) which means that we expect on VNIR focal plane array the hit rates reported in Table 3:

| Focal plane array (pixels) | 256x256 | 512x512 |
|---|---|---|
| Pixel size (µm) | 30 | 18 |
| Mean rate (events/s) | 11 | 15 |
| Events in 600s | 6600 | 9000 |
| Events in 3s | 33 | 45 |

Table 3

Assuming that when a cosmic ray impacts a detector any unrecorded information stored in the focal plane at that location would be lost; thus if five pixel per event were affected by the hit, the data loss would be as follow:

| Focal plane array (pixels) | 256x256 | 512x512 |
|---|---|---|
| Pixel size (µm) | 30 | 18 |
| Pixels affected in 3s | 0.25% | 0.10% |
| Pixels affected in 600s | 50.35% | 17.15% |

Table 4

The main conclusion of our analysis is that two different readout rates and sampling methods are needed for bright sources and faint sources. With the noise performances considered for the Teledyne MCT detectors (H2RG), it is possible to meet the noise requirements well within the maximum allowed integration times in both cases.

The expected cosmic rays hits date, based on studies carried out for JWST, allowed us to estimate that the percentage of pixels that will be affected by the glitches will be very low and that it will not be necessary to correct for the cosmic hits effects in case of bright sources, where it will be sufficient to identify and discard the affected readouts (only a max 0.25% of the overall array will be affected by the cosmic hits in a 3 secs exposure). In case of faint sources a more detailed evaluation will be performed in the future, to assess the real need to implement a deglitching procedure onboard.

**6. The detector's electronics**

The MCT-based detector will be coupled with a ROIC (Read Out Integrated Circuit) bump bonded to the device's sensitive area. The ROIC will act as a proximity electronics in order to extract the low level noise analogue signal from the detector, addressing the very low power dissipation requirements imposed by the environmental thermal aspects (for example, the power dissipation of the Selex detector+ROIC is < 5 mW). The analogue signal will be amplified by the ROIC output OPAMP(s) (typically 4 or 8 for the two detector halves collecting respectively the VIS and NIR signals of the target spectrum) and fed to the cold front-end electronics (CFEE) where A/D conversion will take place. CFEE is connected to the warm section of the payload electronics by means of suitable low thermal conductive harness (Morgante and Terenzi 2013). The payload's warm electronics is essentially constituted by the warm front-end electronics (WFEE) generating driving signals for the detector ROIC/CFEE and the Instrument Control Unit (ICU) acting as the main payload processing electronics and collecting the digitized signals from all scientific channels. WFEEs will reside in a box specifically designed and located near the ICU which will be kept at a temperature in the range 0÷40°C.

The detector is expected to be integrated easily and operate well with a range of electronics solutions. The distance between the Detector Sub Assembly and the CFEE and between the CFEE and the WFEE appears unavoidable in this system presentation and introduces technical challenges associated with a distributed signal chain including driving load capacitance, achieving settling, minimizing cross talk, ensuring stability and reducing noise.

A key challenge will be ensuring that the active differential drive circuit power dissipation can be reduced to an acceptable level for the thermal design constraint and cold electronics operating temperature. The number of video output channels and the operating speed will inform the circuit choice. For a 4-channel circuit operating at 5MHz pixel rate, it is expected that the CFEE active power dissipation could be reduced to around 50mW whereas for an 8-channel circuit operating at 10MHz pixel rate, the expected power dissipation would increase to 200÷300mW. Clearly decreasing the power is desirable and would be a key aim and design trade in the design phase.

The cold front end electronics circuit technology and the level to which the actual power dissipation can be reduced is likely to be a key element to determine whether the overall electronics will only need the cold (45K) section or both cold and warm (120K) sections in the system. It is considered desirable to implement the digital differential drive circuits near to the detector, where possible, to minimize the common mode noise.

The baseline detector can be easily interfaced with the SIDECAR electronics solution, which helps to mitigate a number of electronics design challenges in implementing a full-functional solution. The key benefit is the closer integration of the ADC to the detector which is expected to simplify the interface design, safeguard SNR and mitigate cross talk and some noise sources. Also the Selex detector interface signals are understood to be compatible with the SIDECAR digital and analogue interfaces and the video may be optimized by using gain and offset adjustments.

Functionally, the WFEE and CFEE shall configure and operate the detector and shall provide digitized data to the ICU.

The baseline SIDECAR CFEE will receive a master clock and sync signals to be properly operated and to generate the detector clocks and control signals. The WFEE can also include an FPGA to provide the CFEE digital clocks and serial interfaces. These I/F could be directly integrated inside the EChO ICU but this solution will be analyzed in more detail during the next phase as the WFEEs could be used to perform some digital preprocessing (like digital masking) on images before sending them to the ICU in order to distribute the digital signal processing tasks and data collecting procedures.

The FPGA could also provide a data/HKs multiplexing function to simplify the interfaces to the ICU although separate channels are preferred to improve redundancy at least for scientific data.

The WFEE, if definitively adopted, shall implement stabilized voltage regulators and bias generators for the CFEE and detector circuits and shall interface the CFEE using a suitable cryo harness design (Morgante and Terenzi 2013; Focardi et al. 2014). This critical subsystem is designed as part of the signal interface between the detector, CFEE and WFEE in order to ensure that the best system design, design trades and required signals performance are achieved by design.

## 7. Summary

In the present paper the scientific objectives of the EChO mission have been presented. The VNIR module has been designed to fulfill both technical and scientific requirements of the proposed mission. Some of the adopted technical solutions have been shown.

## 8. Acknowledgments

The authors wish to thank the Italian Space Agency for the financial support to the EChO programme. They are also grateful to the other members of EChO consortium and to the European Space Agency for their support during the phase A study. The present research has been funded by the contract ASI-INAF I/022/12/0.

# References

1) Farina, M., Di Giorgio A. M., Focardi, M.: Noise evaluation for ECHO VNIR detectors. http://sci.esa.int/science-e/www/object/doc.cfm?fobjectid=53425 (2013). Accessed 3 January 2014

2) Finger, G., Dorn, R.J. , Meyer, M., Mehrgan, L., Stegmeier, J., Moorwood, A., " Performance and Evaluation of Large Format 2 k x 2 k MBE Grown MCT HAWAII-2RG Arrays Operating in 32-channel mode",  Proc. of "Workshop on Scientific Detectors for Astronomy, Taormina (Sicily/Italy), June 20 - 24, 2005", eds. P. Amico and J.W. Beletic, Springer Verlag (2005)

3) Fixsen, J., Offenberg, J.D., et al., "Cosmic-Ray rejection and Readout Efficiency for Large-Area Arrays", PASP 112, 1350-1359 (2000)

4) Focardi, M., Farina, M., Pancrazzi, M., Di Giorgio, A. M., Pezzuto, S., Ottensamer, R., Pace, E., Micela, G.: EChO electronics architecture and SW design. Experimental Astronomy EChO Special Issue, 2014

5) Morgante, G., Terenzi, L. : TMM/GMM Description and Results. http://sci.esa.int/science-e/www/object/doc.cfm?fobjectid=53425 (2013). Accessed 7 January 2014

6) Rauscher, B. J., Isaacs, J.C., "Cosmic Ray Management on NGST 1: The effect of Cosmic Rays on Near Infrared Imaging Exposure Time",  STScI-NGST-R-0003A (2000)

7) Rauscher, B. J., Fox, O., Ferruit, P. et al." Detectors for the James Webb Space Telescope Near-Infrared Spectrograph I: Readout Mode, Noise Model, and Calibration Considerations", PASP, 119, 768 (2007)

8) The Extrasolar Planet Encyclopaedia http://www.exoplanet.eu/ (2014). Accessed 9 January 2014.

9) Tinetti et al: EChO. Exoplanet characterisation observatory, Experimental Astronomy, Volume 34, Issue 2, 311-353 (2012)

10) Waldmann, I. and Pascale E.: EChO Pointing Jitter Impact on Photometric Stability. http://sci.esa.int/science-e/www/object/doc.cfm?fobjectid=53425 (2013). Accessed 3 January 2014

**Figure captions**

**Fig. 1** Optical layout of the VNIR spectrometer

**Fig. 2** Grating diffraction orders projected on the VNIR detector, starting from m = 3 at the bottom (near infrared spectral range) to m = 20 on the top (visual spectral range). The wavelengths in nanometers are also indicated

**Fig. 3** Spectrum of the internal calibration lamp used to check the instrument stability

**Fig. 4** Sketch of the internal calibration unit. The calibration unit is equipped with two halogen-tungsten lamps for redundancy. The lamps inject their light into an integration sphere, having two output fibers that will feed the two input fibers to the spectrometer (ranges 0.4-1.0 µm and 1.0-2.5 µm respectively). The calibration unit is located in a separate box on a side of the service box where the mirrors collect the light from the VNIR feeding optics and focus it on the optical fibers inputs

**Fig. 5** VNIR box close (left panels), open (upper right panel) without the internal calibration unit and top view of the spectrometer elements (lower right panel). The path of the light inside the instrument is shown in green

**Fig. 6** VNIR efficiency: black curve, present estimation (this study); red curve: estimation done by using coated aluminium for improving the relative efficiency below 1 µm; blue curve, estimation done by using protected silver for all the mirrors

**Fig. 7** Normalized energy loss vs MPE at 0.8 µm (dashed line) and 2.5 µm (solid line)

**Fig. 8** Photometric error induce by the combine RPE/PRE at 0.8 µm (left) and 2.5 µm (right) in one second of integration. The solid lines correspond to the RPE1 (black), RPE2 (red) and RPE3 (green) cases discussed in the text

**Fig. 9** Spot diagram of focused system in 50 micron at central field 0.0° in blue and marginal field 0.00000278° in green colour (on left side) and Spot diagram of defocused system in 50 µm at central field 0.0° in blue and marginal field 0.00000278° in green colour (right side)

**Fig. 10** Non destructive MCT detectors readout sampling scheme

**Tables**

**Table 1**: Main EChO VNIR module characteristics

**Table 2**: Resuming data of Encircled Energy at focused and defocused system (WFE 250 rms)

**Table 3**: Expected cosmic rays hits rate for VNIR focal plane array positioned in lagrangian point L2

**Table 4:** Expected data loss, assuming 5 pixels affected per hit

# Figures

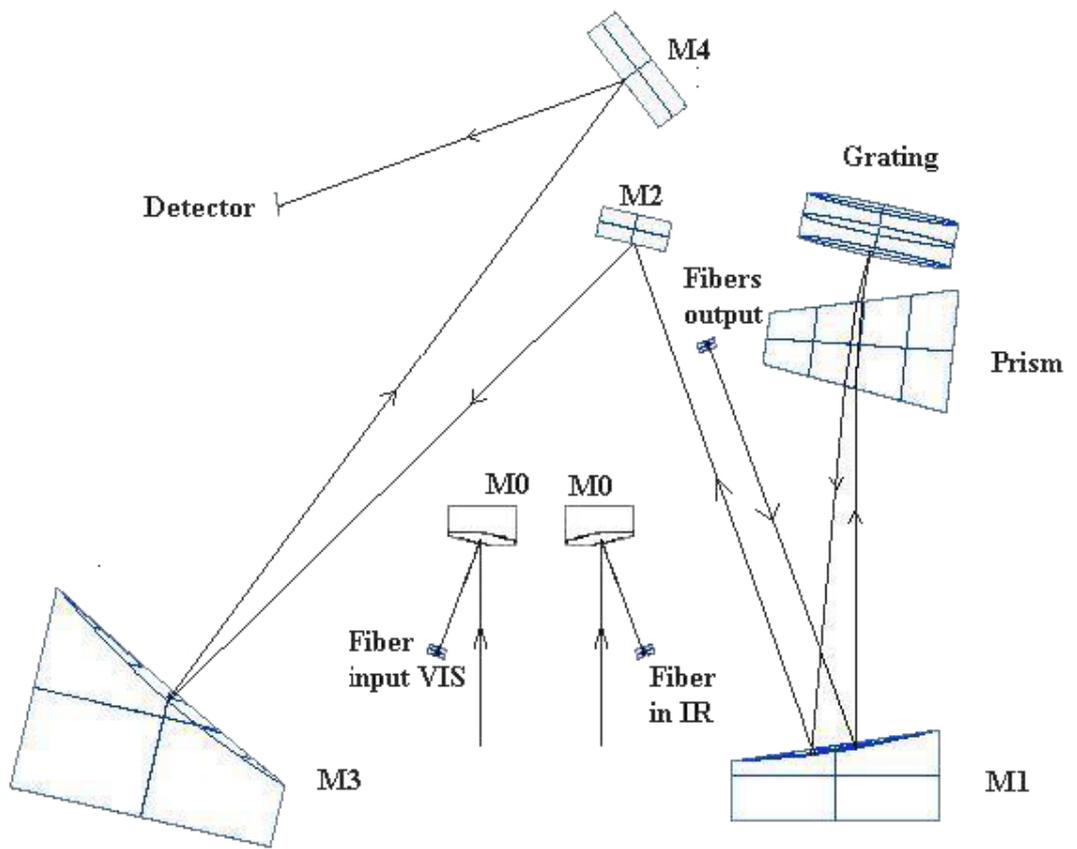

Fig. 1

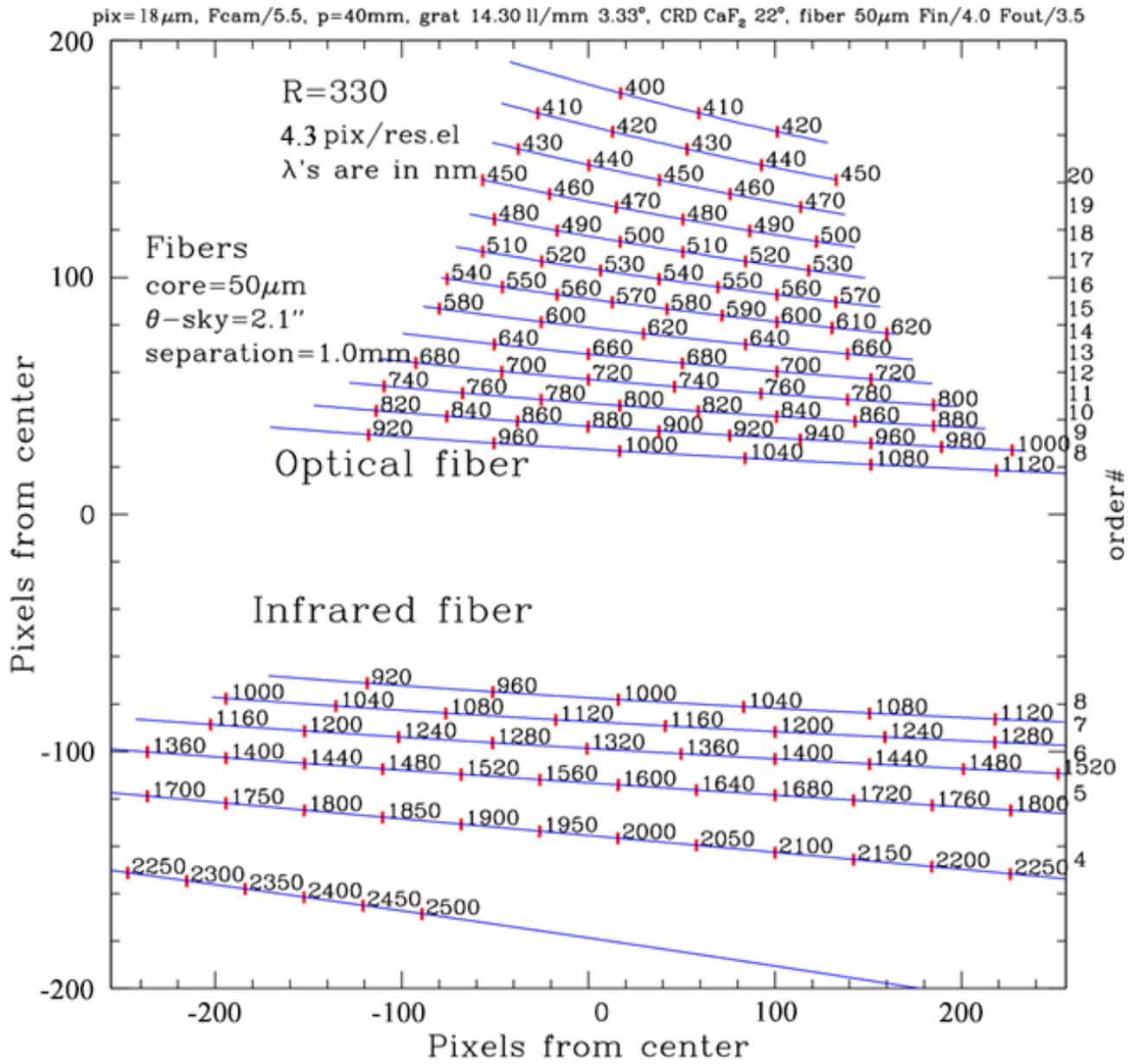

Fig. 2

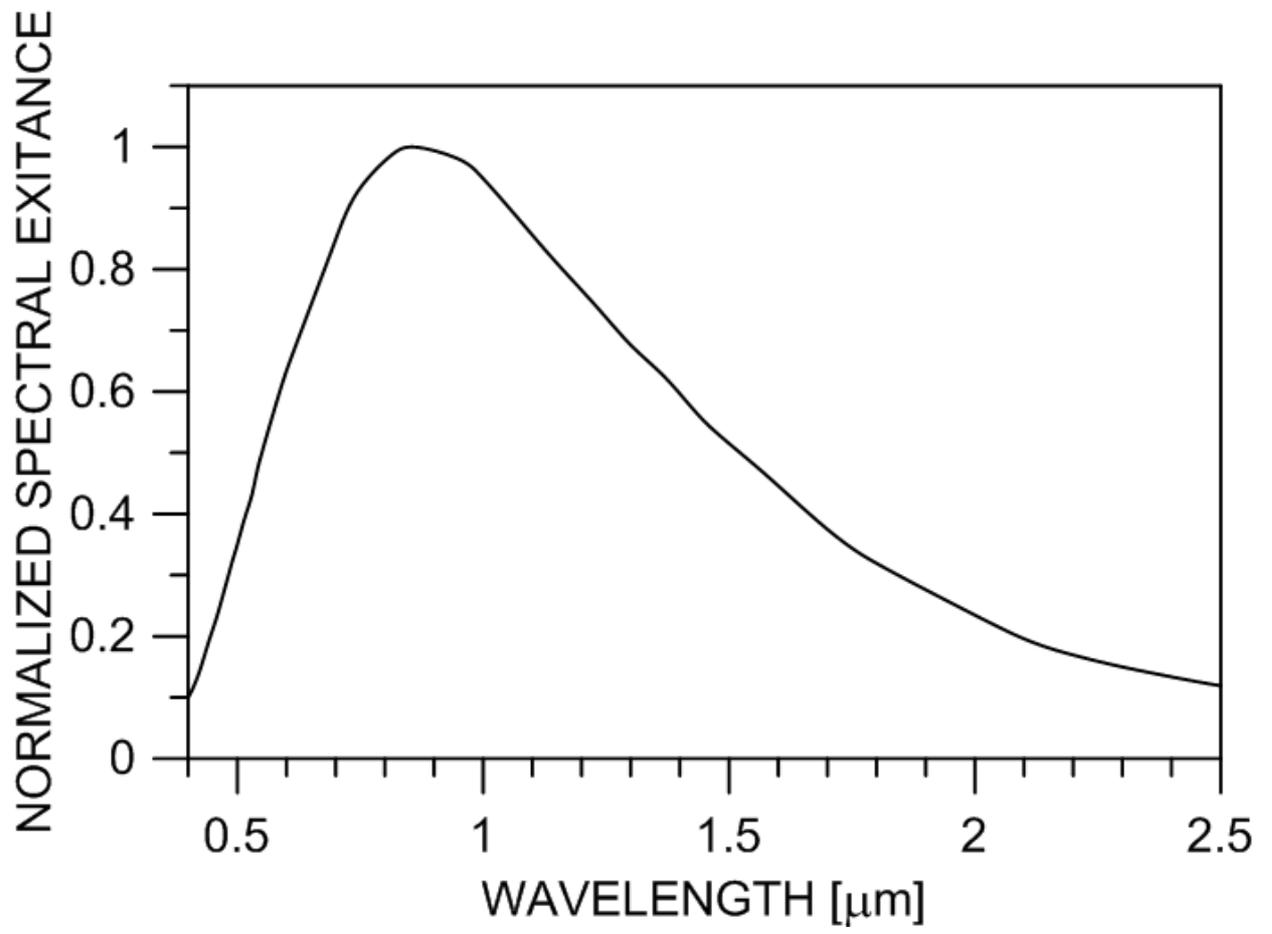

Fig. 3

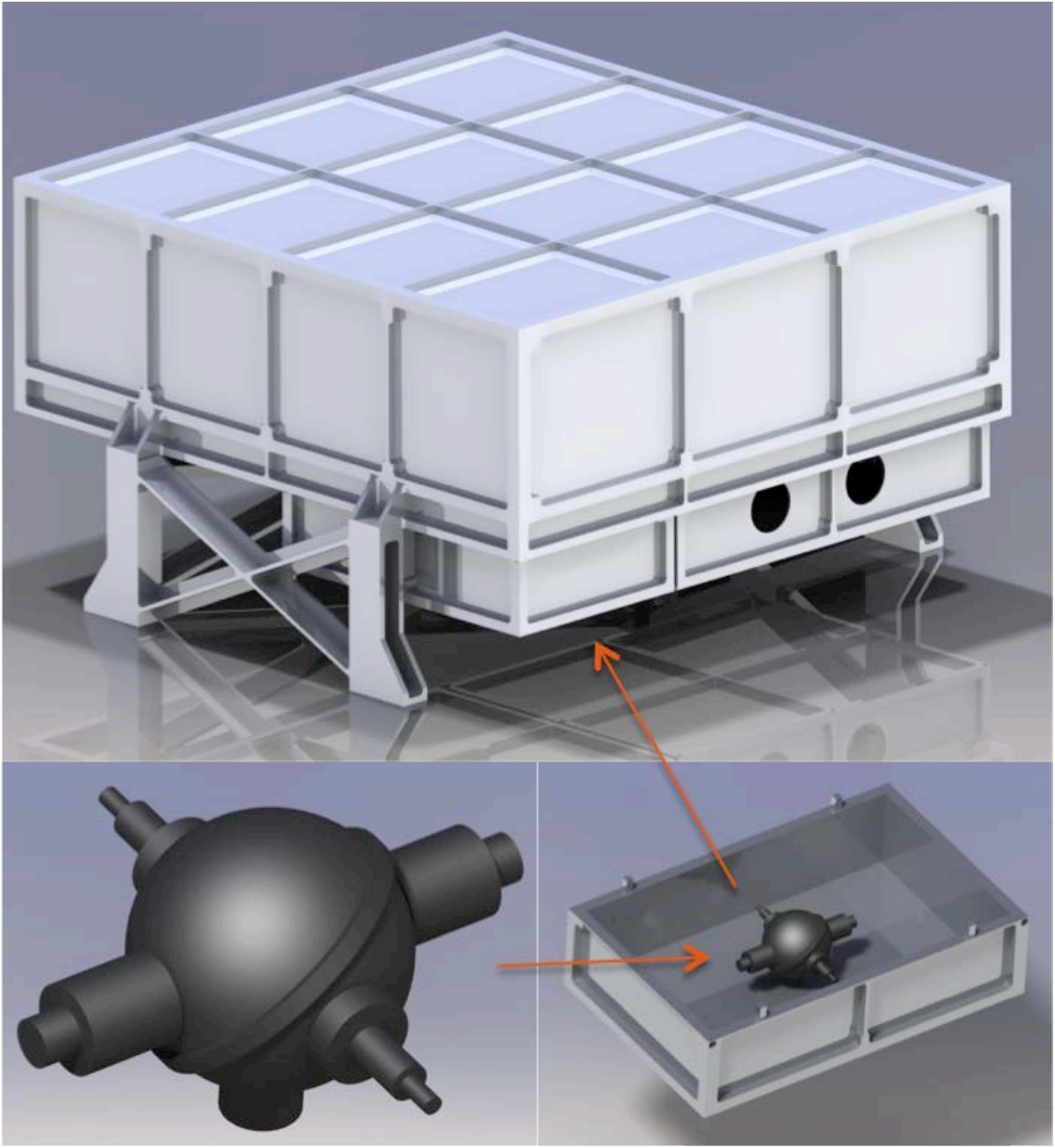

Fig. 4

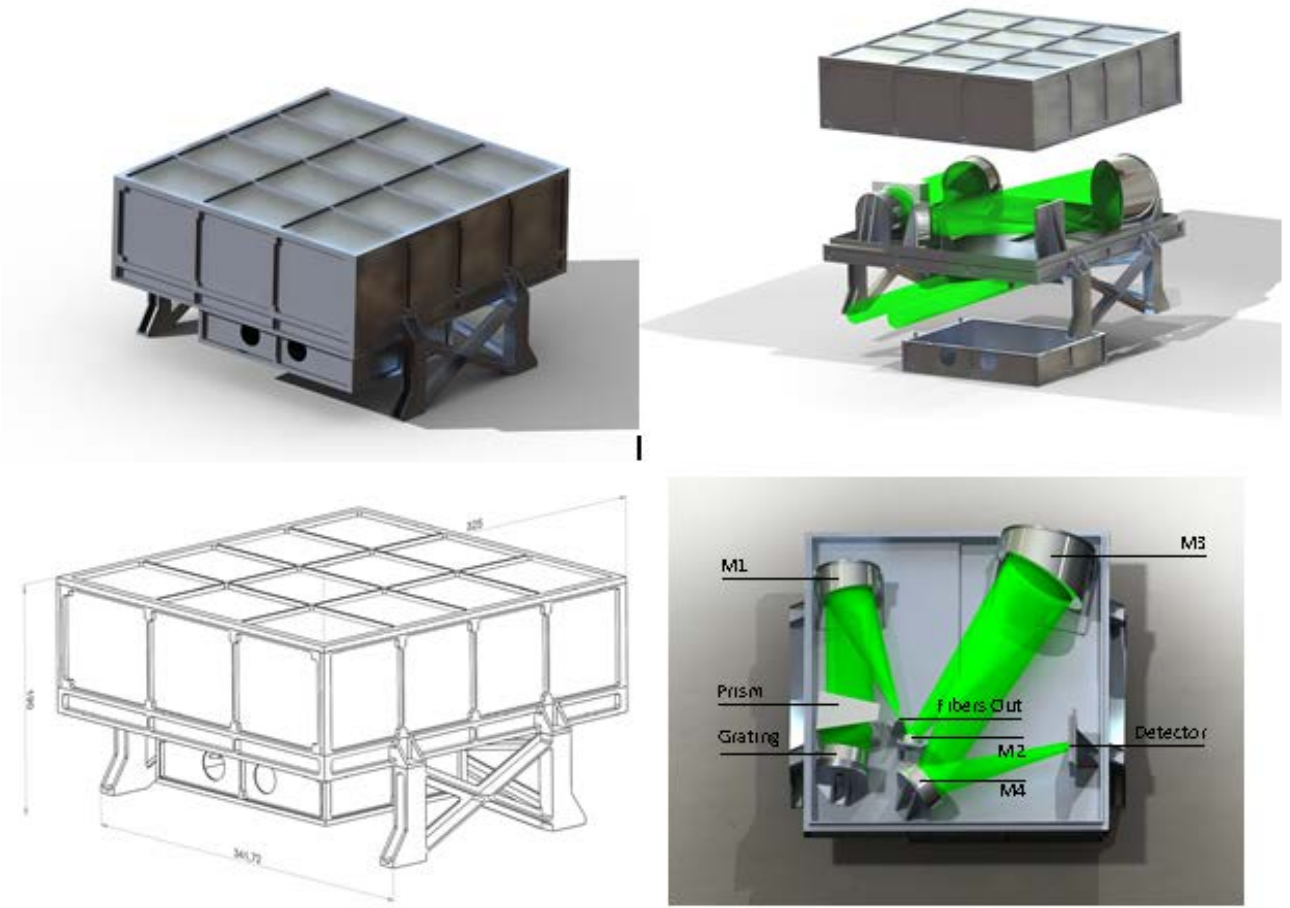

Fig. 5

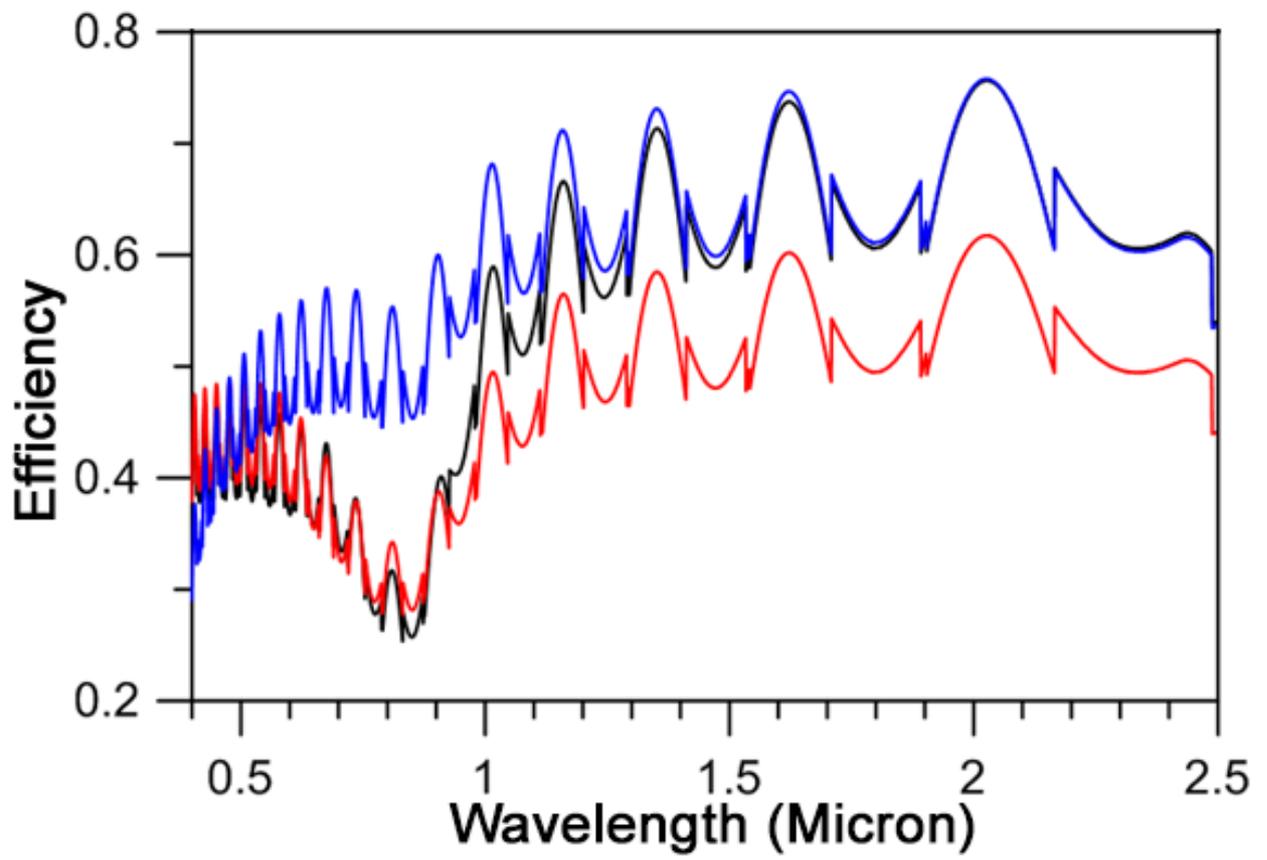

Fig. 6

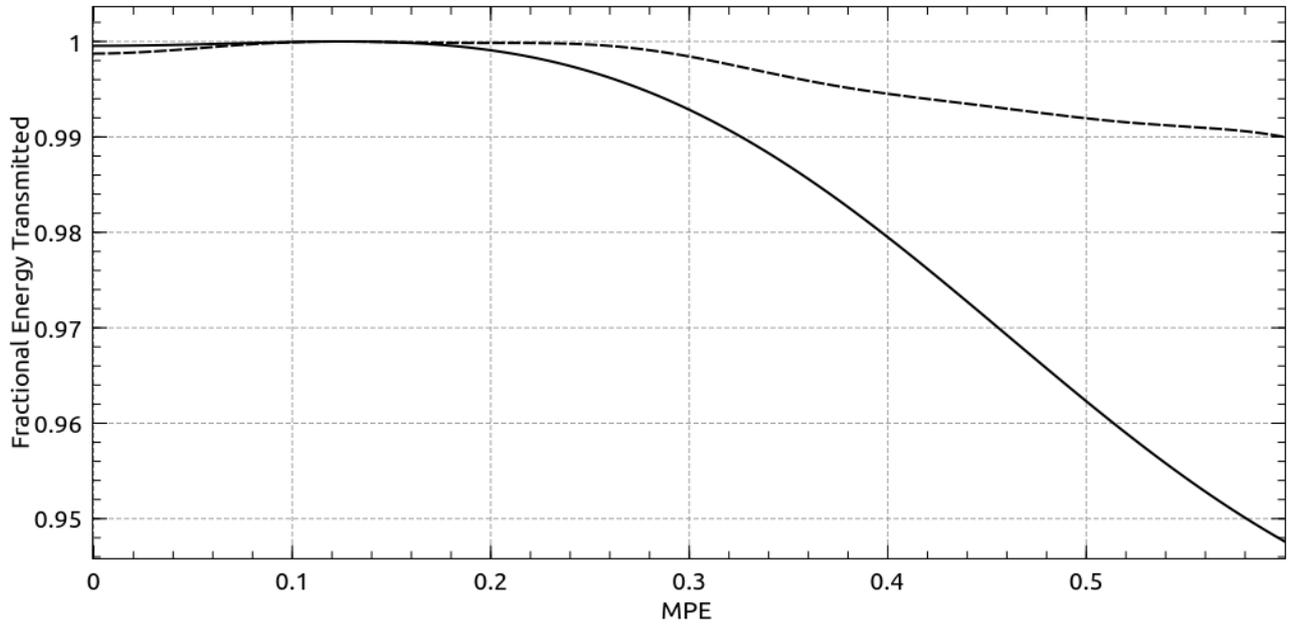

Fig. 7

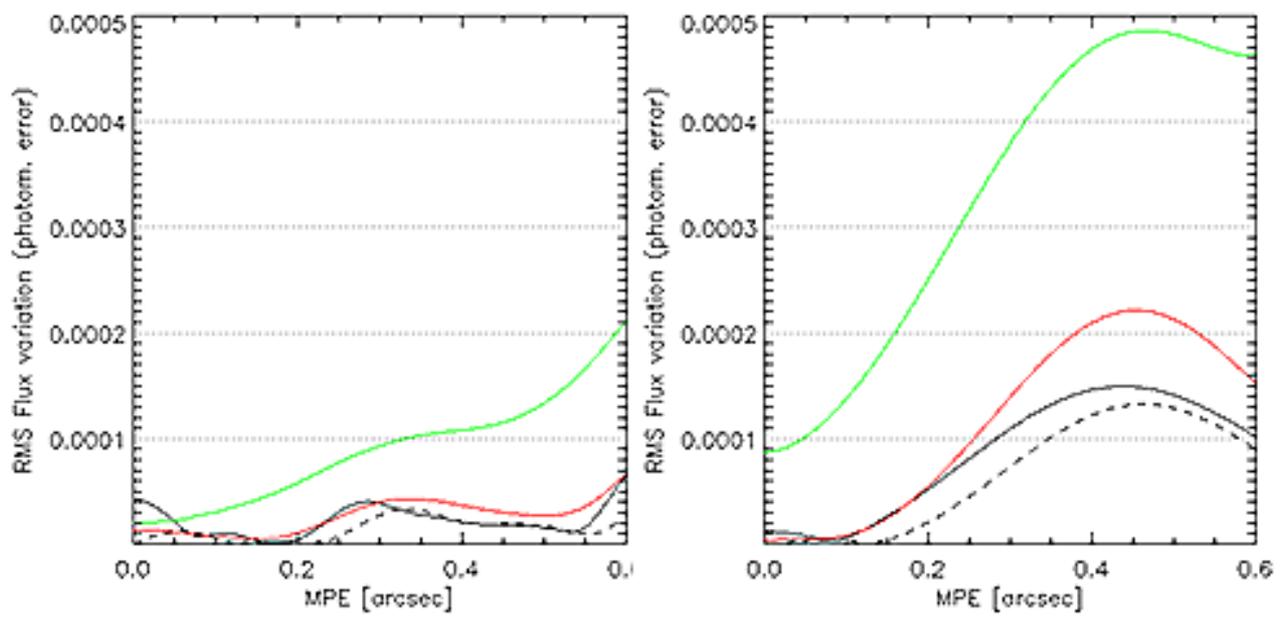

Fig. 8

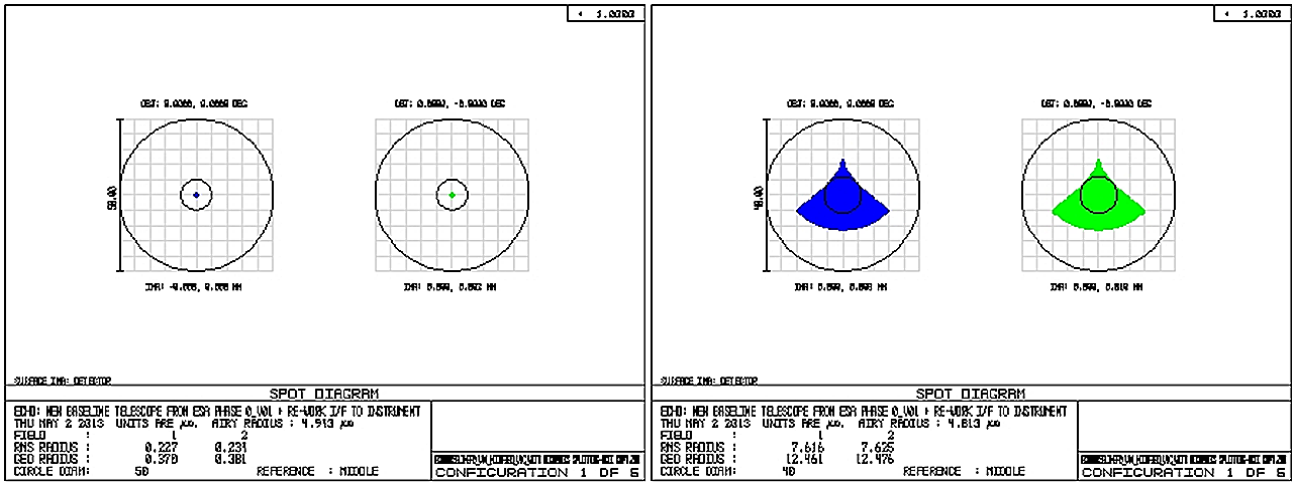

Fig. 9

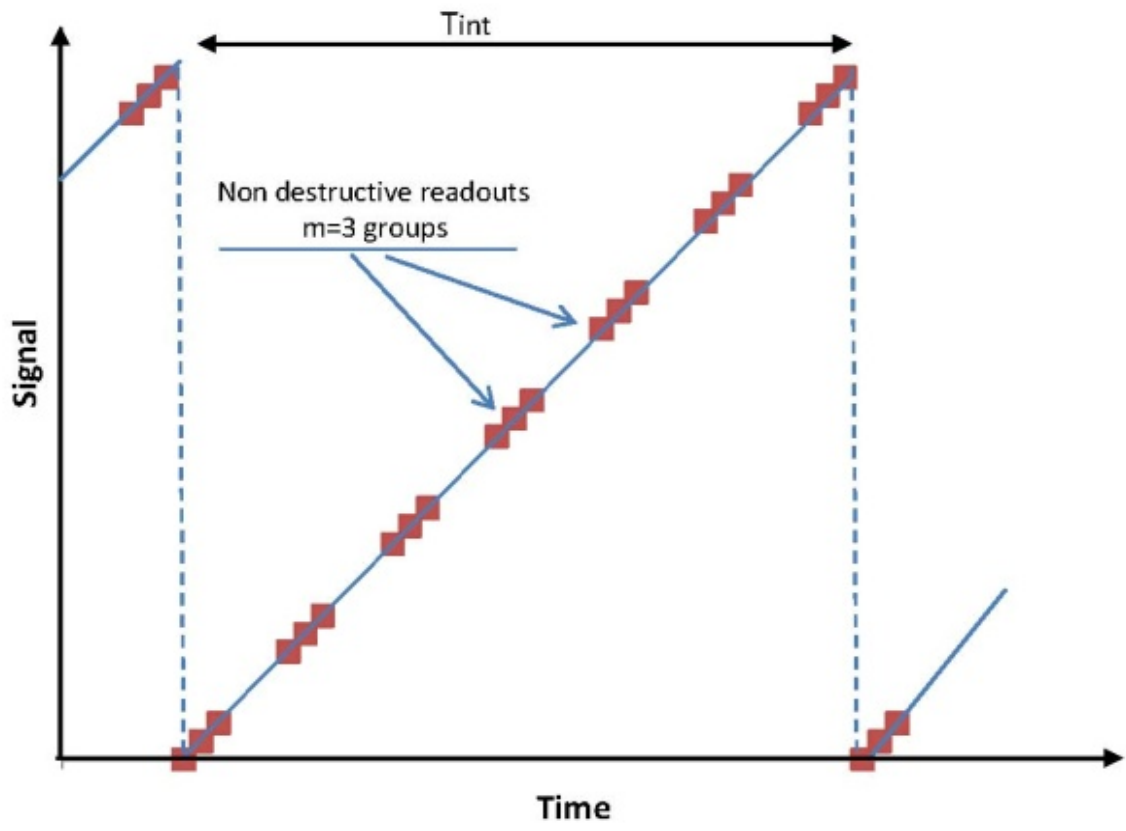

Fig. 10